\documentclass[letterpaper,twocolumn,twoside]{IEEEtran}
\usepackage{amsmath, amssymb, bm, cite, epsfig, psfrag}
\usepackage{multirow}

\def\beq{\begin{equation}}
\def\eeq{\end{equation}}
\def\beqa{\begin{eqnarray}}
\def\eeqa{\end{eqnarray}}
\def\beqan{\begin{eqnarray*}}
\def\eeqan{\end{eqnarray*}}

\def\R{{\mathbb{R}}}
\def\C{{\mathbb{C}}}

\def\argmax{\mathop{\mathrm{arg\,max}}}

\def\x{\times}

\def\la{\leftarrow}
\def\ra{\rightarrow}

\def\arr{\rightarrow}
\def\Exp{\mathbf{E}}

\def\Xset{{\cal X}}

\newcommand{\bbf}{\mathbf{b}}

\newcommand{\gbf}{\mathbf{g}}

\newcommand{\sbfhat}{\widehat{\mathbf{s}}}
\newcommand{\ubf}{\mathbf{u}}

\newcommand{\vbfhat}{\widehat{\mathbf{v}}}

\newcommand{\xbf}{\mathbf{x}}
\newcommand{\xbfhat}{\widehat{\mathbf{x}}}

\newcommand{\zbf}{\mathbf{z}}
\newcommand{\zbfhat}{\widehat{\mathbf{z}}}
\newcommand{\Abf}{\mathbf{A}}

\newcommand{\Qbf}{\mathbf{Q}}
\newcommand{\Rbar}{\overline{R}}

\def\Nrx{N_{\rm rx}}
\def\Ntx{N_{\rm tx}}

\title{Belief Propagation Methods for Intercell Interference Coordination}
\author{Sundeep Rangan, ~\IEEEmembership{Member,~IEEE,}
        Ritesh \ Madan,~\IEEEmembership{Member,~IEEE}
\thanks{S.\ Rangan (email: srangan@poly.edu) is with
        Polytechnic Institute of New York University, Brooklyn, NY.}
\thanks{R.\ Madan (email: rmadan@qualcomm.com) is with
        Qualcomm, Bridgewater, NJ.}
}

\begin{document}


\setcounter{page}{1}

\maketitle
\begin{abstract}
We consider a broad class of interference coordination and resource
allocation problems for wireless links where the goal is to maximize
the sum of functions of individual link rates. Such problems arise
in the context of, for example, fractional frequency reuse (FFR) for
macro-cellular networks and dynamic interference management in
femtocells. The resulting optimization problems are typically hard
to solve optimally even using centralized algorithms but are an
essential computational step in implementing rate-fair and queue
stabilizing scheduling policies in wireless networks. We consider a
belief propagation framework to solve such problems approximately.
In particular, we construct approximations to the belief propagation
iterations to obtain computationally simple and distributed
algorithms with low communication overhead. Notably, our methods are
very general and apply to, for example, the optimization of transmit
powers, transmit beamforming vectors, and sub-band allocation to
maximize the above objective. Numerical results for femtocell
deployments demonstrate that such algorithms compute a very good
operating point in typically just a couple of iterations.
\end{abstract}



\section{Introduction}
Interference coordination has re-emerged as a fundamental challenge
for next-generation cellular wireless systems. Traditional
macrocellular deployments are likely to be supplemented with smaller
femtocells and relays, with mixtures of restricted and open access,
often deployed in an ad hoc manner \cite{ChaAndG:08,LopezVRZ:09}.
Such deployments may create much stronger and highly variable (in
time and space) interference conditions than those experienced in
current macrocellular networks, and traditional cellular power and
rate control may not be adequate \cite{FemtoForum:10}. To address
this challenge, a key focus of the current 3GPP LTE-Advanced
standardization efforts is on the design of an interference
coordination framework for such unplanned cellular deployments of
base-stations with widely different transmission
powers~\cite{3GPPICIC}. Current release of the LTE
specification~\cite{3GPP36.300} provides simple methods for inter-cell
interference coordination (ICIC), see e.g.,~\cite{fodor_2009}. Along
with the design of mechanisms for ICIC, algorithms to exploit such
mechanisms are an active area of research as well, see for
example,~\cite{kang_2008},~\cite{sundaresan_2008}.

Mathematically, interference coordination is a complex distributed
optimization problem involving scheduling decisions at the
transmitters of multiple interfering links.
In this work, we consider a general \emph{linear mixing}
interference model where the scheduling decisions
in each link are represented as vector (e.g., transmission
powers on different sub-bands in frequency, beamforming weights) and
the interference on each link is a linear
combination of the scheduling vectors on the other links.
Associated with each link at a given time is a \emph{utility function} which
describes the benefit to a link as a function of the scheduling
vector from the serving transmitter and interference from the other
transmitters.  The linear mixing model is extremely general
and can apply to a large class of interference models and objectives.
Computing maximum weighted matching for queue
stability~\cite{TassiulasE:92} and maximization of sum utility of
average rates for fairness~\cite{stolyar_2005} are special cases of
this formulation.

In the past few years, algorithms for special cases of the above
problem have been extensively studied. In many cases, algorithms
with provable desired properties have been obtained -- for example,
there is a rich literature on distributed power control methods to
achieve a desired SINR for each link~\cite{chiang_2008} and to
maximize a certain class of utility functions of
SINRs~\cite{huang_2006}, approximation algorithms for maximum weight
matching for combinatorial interference model were obtained
in~\cite{sharma_2007, gupta_2007}, efficient methods to compute
optimal beamforming vector for multiuser downlink~\cite{TanCS:2010}
and maximizing sum utility of SINRs on uplink~\cite{tan_isit},
stabilizing policies for collision sense multiple access (CSMA) type
of models based on simulated annealing were derived
in~\cite{rajagopalan_2009}. More generally, heuristic algorithms
have been constructed to solve certain specific problems
approximately in, for example,~\cite{StolyarVis:2009,madan_2010}.
While these algorithms perform well in practice in spite of no
provable guarantees, the insights and approximations used to obtain
these algorithms are very specific to the problem under
consideration.

In this paper, we make the following contributions:

\noindent\emph{BP Framework:} We consider a belief propagation (BP)
framework for a very
general wireless scheduling and interference coordination
optimization problem. The underlying optimization problem is posed
as a problem of estimating marginals of a joint probability
distribution. BP provides a systematic and general approach to
obtain distributed algorithms; it can be used with arbitrary
nonlinear utility functions and scheduling vectors sets, which
enable the algorithm to be applied a range of complex scheduling
problems including power control, subband scheduling and distributed
beamforming.
Also, while we do not obtain any theoretical guarantees, in
practice a few iterations of BP generate a good operating point.
Thus, it typically has faster convergence than gradient based
algorithms (e.g.,10s of iterations in~\cite{tan_isit}) or simulated
annealing~\cite{Hajek:88}.

\noindent \emph{Approximation Algorithms:} It is well known that
implementing BP for distributed optimization problems entails high
computational complexity and communication overhead.
Exploiting the linear mixing interference model and applying
Gaussian  and first-order approximations
similar to~\cite{BouCaire:02,MontanariT:06,GuoW:06,BayatiM:10arxiv,Rangan:10arXiv},
we develop an approximate BP method that has low complexity,
distributed implementation and minimal messaging.
Along many links, messages can
be carried in small payloads and can be broadcast without separate
unicast transmissions, which is particularly crucial for wireless
systems.  Moreover, the resulting algorithm has a natural interpretation
has a ``soft" RTS / CTS type handshaking.  The approximate BP algorithm
is also similar to the recent approximate message passing (AMP) algorithm
in \cite{DonohoMM:09arxiv,BayatiM:10arxiv} and this connection may be useful
for further analysis.

\noindent \emph{Numerical Results:} Through simulations for
femtocell deployments, we demonstrate that approximate BP provides
good performance for sub-band and power allocation to maximize
utilities of rates, sub-band and power allocation to maximize a
weighted sum of rates, and beamforming optimization to maximize
utilities of average rates. Also, although the BP algorithm requires
multiple exchange of messages before each scheduling decision, our
simulations indicate good performance with only two rounds of
messaging. Thus, the approximate BP approach is a promising paradigm
for new emerging cellular deployments with large interference
variations in time and space compared to current predominantly
macro-only deployments.

\subsection{Previous Work}

Algorithms for ICIC in LTE macrocells have been considered in a
large number of works in both the uplink and downlink
\cite{RahmanYW:09,FodorKRRSM:09,XiangLH:07}. These works are
generally based on adaptive subband scheduling and
fractional frequency reuse (FFR) methods
\cite{HanPLAJ:08,StolyarVis:2009} and exploit statistics over large
numbers of mobiles per macrocell.
Interference mitigation in femtocells has focussed on similar techniques as well
as frequency planning, power control~\cite{chandra_2008}, or
semi-static resource allocation~\cite{kang_2008,humblet_2009,sundaresan_2008}.
As we will demonstrate
in the simulations, the BP methods presented here can also be used
for adaptive subband scheduling as a special case of the linear mixing interference
model.
Also, much of the ICIC work has considered slowly varying allocations
that don't change over few 10s to 100s of milliseconds.
Due to the low messaging overhead, it is possible that approximate BP
can also be used for more dynamic interference
management in femtocell deployments, where there is high variability
in load and interference from one timeslot to another.

For scheduling based on more dynamic traffic statistics such as
queue lengths and head-of-line delays, variants of maximum weight
scheduling can be used~\cite{Tassiulas:98},\cite{shakkottai_2001}.
Unfortunately, computing a maximum weight schedule is generally
NP-hard, and much work has thus focused on approximate algorithms.
In addition to the works mentioned above,
the works \cite{Tassiulas:98} and \cite{GiacconePS:03} proposed
randomized linear complexity (but centralized) algorithms, and
\cite{JiangWal:08,RajagopalanSS:09,NiSrikant:09} present simple distributed
algorithms for combinatorial interference models. Greedy maximal
weight matching for such interference models has been considered
in~\cite{DimakisW:06,JooLinShroff:09,LeconteNS:09}.

However, many of the above works apply a hard constraint
interference models, where neighboring links cannot transmit
simultaneously. Cellular systems in contrast permit multiple
interfering links to transmit simultaneously and then use rate
control to adapt to the resulting signal-to-interference and noise
ratio (SINR). Thus, the degradation in rate with interference is
gradual, and are difficult to capture in the combinatorial
interference model. In contrast, ``soft" interference effects can be
easily modeled in the BP utility framework.

However, we note that there is an important theoretical connection
between the methods in \cite{JiangWal:08,RajagopalanSS:09} and
the BP method considered here. As we will discuss below, BP arrives
at the scheduling decision by estimating the marginals of a certain
joint probability distribution function given in \eqref{eq:probF}.
The CSMA-type methods in \cite{JiangWal:08,RajagopalanSS:09}
can be seen as a simulated annealing
(SA) method for selecting a random scheduling vector from precisely
the same distribution in the context of a constraint combinatorial
interference model. SA can be seen as an asymptotically
exact but slow method \cite{Hajek:88} for solving the optimization problem.
In constrast, BP is approximate, but potentially faster.

General overviews of BP can be found in a number of works including
\cite{Pearl:88,WainwrightJ:08}. In the context of wireless
scheduling, theoretical guarantees have been obtained for on-off
channels and the combinatorial contention graph
model~\cite{SanghaviMW:07, BayatiSS:08}. The methods here can be
seen as a generalization of these methods to soft interference
models with larger class of scheduling vectors.

\section{Problem Formulation} \label{sec:prob}

We consider a wireless scheduling problem with $n$ links.
The transmitter of each link $j=1,\ldots,n$, denoted TX $j$,
must select some \emph{scheduling vector}
$\xbf_j \in \Xset \subseteq \R^{n_x}$, which contains
$n_x$ parameters related to link $i$.
Examples of these parameters will be given below.
The selection of the scheduling vectors results in
an \emph{interference vector} $\zbf_i \in \R^{n_z}$
at the receiver of each link $i$, denoted RX $i$.
The interference is assumed to be
a linear function of the scheduling vectors of the other links,
\beq \label{eq:zidef}
    \zbf_i = \sum_{j=1}^n \Abf_{ij}\xbf_j
\eeq
for some matrices $\Abf_{ij} \in \R^{n_z \times n_x}$.  We assume that
$\Abf_{ii} = 0$, so that link $i$ does not interfere with itself.
We will let $\xbf$ and $\zbf$ be the column vectors with entries
$\xbf_j$ and $\zbf_i$,
\beqan
    \xbf &=& \left[ \xbf_1' \cdots \xbf_n'\right]' \in \R^{n_xn} \\
    \zbf &=& \left[ \zbf_1' \cdots \zbf_n'\right]' \in \R^{n_zn},
\eeqan
and write $\zbf = \Abf \xbf$ where $\Abf$ is the block matrix with
entries $\Abf_{ij}$. We call $\Abf$ the \emph{interference matrix}.
Also, we let $\Abf_i$ denote the $i$th column of the $\Abf$ so that
$\zbf_i = \Abf_i\xbf$.

Associated with each link $i$, is some utility function
$f_i(\xbf_i,\zbf_i)$ of the scheduling vector $\xbf_i$
and interference vector $\zbf_i$.  The scheduling problem is to
maximize the overall utility
\beq \label{eq:opt}
    \max_{\xbf} F(\xbf), \ \ \ F(\xbf) = \sum_{i=1}^n f_i(\xbf_i, \zbf_i).
\eeq
We will sometimes call the optimization problem \eqref{eq:opt},
an optimization with \emph{linear mixing} to stress the linear dependence
of the interference on the transmit vectors.

\section{Linear Mixing Utility Examples} \label{sec:ex}
The linear mixing formulation above is extremely general
and can incorporate a large class of utility functions and
interference models.

A general treatment of utility functions for wireless scheduling
can be found in \cite{KelleyMT:98,ShakkottaiS:07}.
In our simulations, we will consider utility maximization for both
static and time-varying problems.
For static optimization, the scheduling vectors $\xbf_j$
are selected once for a long time period
and the utility function is typically of the form
\beq \label{eq:utilStatic}
    f_i(\xbf_i,\zbf_i) = U_i(R_i(\xbf_i,\zbf_i)),
\eeq
where $R_i(\xbf_i,\zbf_i)$ is the long-term rate as a function of the
TX vector $\xbf_i$ and interference $\zbf_i$ and $U_i(R)$ is the utility
as a function of the rate.
The problem formulation above can incorporate any of the
common utility functions including: $U_i(R) = R$ which
results in a sum rate optimization; $U_i(R) = \log(R)$ which is the proportional
fair metric and $U_i(R) = -\beta R^{-\beta}$ for some $\beta > 0$ called
an $\beta$-fair utility.  Penalties can also be added if there is a cost
associated with the selection of the TX vector $\xbf_j$ such as power.

To accommodate time-varying channels and traffic loads,
many cellular systems enable fast dynamic scheduling in time slots
in the order of 1 to 2 ms.  For these systems, the utility maximization can be
re-run in each time slot.  One common approach is that
in each time slot $t=0,1,2\ldots$,
the scheduler uses a utility of the form
\beq \label{eq:wtUtil}
   f_i(t,\xbf_i,\zbf_i) = w_i(t)R_i(t,\xbf_i,\zbf_i), \ \ \
\eeq
where $w_i(t)$ is a time-varying weight given by the marginal utility
\beq \label{eq:wtMargUtil}
   w_i(t) = \frac{\partial U_i(\Rbar_i(t))}{\partial R},
\eeq and $\Rbar_i(t)$ is exponentially weighted average rate updated
as \beq \label{eq:rateFilt}
    \Rbar_i(t+1) = (1-\alpha)\Rbar_i(t) + \alpha
    R_i(t,\xbfhat_i(t),\zbfhat_i(t)),
\eeq
where $\xbfhat_i(t)$ is the TX vector and  $ \zbfhat_i(t)$ is
the interference for link $i$ at time $t$. Any maxima of the
optimization \eqref{eq:opt} with the weighted utility
\eqref{eq:wtUtil} is called a \emph{maximal weight matching}. A
well-known result of stochastic approximation~\cite{stolyar_2005} is
that if $\alpha \arr 0$, and the scheduler performs the maximum
weight matching with the marginal utilities \eqref{eq:wtMargUtil},
then for a large class of processes, the resulting average rates
will maximize the total utility $\sum_i U_i(\Rbar_i(t))$.

The above utilities are designed for infinite backlog queues.
For delay sensitive traffic, one can take the weights $w_i(t)$ to be the queue length
or head-of-line delay.  Maximal weight matching performed with these weights
generally results in so-called throughput optimal performance \cite{TassiulasE:92}.
These results also apply to multihop networks with the so-called backpressure
weights.

In addition to incorporating general utilities, an appealing feature of the
linear mixing framework is that a large class of interference models
can also be considered, including, for example:
\begin{itemize}
\item \emph{Flat fading with power control:}  In this case, $x_j$ is a scalar representing
the transmit power, and $A_{ij}$ is the gain from TX $j$ to RX $i$, so that $z_i$ is the
total interference at RX $i$.  The rate, $R_i(x_i,z_i)$ can then be described
as a function of the SINR $g_ix_i/z_i$, where $g_i$ is the channel gain along
link $i$.  Arbitrary SINR to rate mappings may be used.
Note that a special case of on-off channels where $x_j$ is zero or a maximum
transmit power can be used.

\item \emph{Multiple subbands:}  The above example is easily extended to the
case of multiple subbands.  As described in the Introduction, subband scheduling
is one of the key motivating features of LTE, but the optimization is difficult.
To handle multiple subbands, we simply let $\xbf_j$ and $\zbf_i$ be the vectors
of transmit  and interference powers in each subband
and $\Abf_{ij}$ be a diagonal matrix with channel gains in each subband.

\item \emph{Beamforming and linear precoding:}  The linear mixing formulation
can also incorporate problems with transmit beamforming or linear
precoding.  For example, suppose a link has
$N$ transmit antennas and one receive antenna.
If each transmitter TX $j$ uses a beamforming vector $\bbf_j \in \C^N$,
and $\gbf_{ij} \in \C^N$ is the channel from TX $j$ to RX $i$,
the interference at RX $i$ is given by
\[
    z_i = \sum_{j \neq i} \gbf_{ij}'\bbf_j\bbf_j'\gbf_{ij},
\]
which is linear in the rank one matrices $\bbf_j\bbf_j'$.
Hence, if we let $\xbf_j \in \C^{N^2}$ be the column vector with entries
of the matrix $\bbf_j\bbf_j'$, the interference $z_i$ can be represented
as a linear combination of the vectors $\xbf_j$.
The idea can also be generalized to precoding matrices with
multiple transmit streams.
\end{itemize}

\section{Belief Propagation} \label{sec:BpAlgo}

\subsection{Standard BP} \label{sec:stdBP}
We begin by briefly reviewing how we would apply standard BP
 to the optimization \eqref{eq:opt}.
Let $u > 0$ and define the probability distribution \beq
\label{eq:probF}
    p(\xbf) = \frac{1}{Z} \exp(uF(\xbf)) = \frac{1}{Z} \prod_{i=1}^n
        \exp(uf_i(\xbf_i,\zbf_i)),
\eeq where $Z$ is a normalization constant called the partition
function (it is a function of $u$).  BP can be seen as a method to
estimate the marginal distributions of the distribution $p(\xbf)$
with respect to the variables $\xbf_j$. From these marginals, one
can compute the marginal expectations $\xbfhat_j = \Exp\left( \xbf_j
\right)$. A standard result of large deviations~\cite{DemboZ:98}
 is that as $u \arr
\infty$,
 under suitable conditions, $p(\xbf)$ concentrates around the
maxima of $F(\xbf)$ and
\[
    \lim_{u \arr \infty} \xbfhat = \argmax_{\xbf} F(\xbf).
\]
So, if we can estimate the marginal expectations of the probability
distribution (\ref{eq:probF}) for large $u$, we can recover a good
estimate for the maximization of (\ref{eq:opt}).

To compute the marginal distributions, BP associates with the
interference matrix $\Abf$ a bipartite graph $G=(V,E)$ called the
\emph{factor} or \emph{Tanner} graph.  The vertices $V$ consists of
$n$ \emph{transmitter nodes} associated with the transmitters TX
$j$, and $n$ \emph{receiver nodes} associated with the receivers RX
$i$. There is an edge $(i,j) \in E$ if and only if $i=j$ or
$\Abf_{ij}$ is non-zero -- that is TX $j$ has some influence on the
interference or signal at RX $i$. We let $\Nrx(i)$ and $\Ntx(j)$ be
the neighbors sets of the nodes RX $i$ and TX $j$ in graph $G$,
respectively.

With this graph, BP iteratively passes \emph{beliefs} along the
edges of the graph that represent estimates of the marginal
distributions of $p(\xbf)$ with respect to the variables $\xbf_j$.
In the context of the wireless scheduling problem, we can interpret
the iterations as \emph{rounds}, where computations are first
performed at the receivers and then at the transmitters. We index
the round by $t$, and let $p_{i \ra j}(t,\cdot): \Xset \mapsto\R$
denote the belief message from RX $i$ to TX $j$ in the receiver half
of the round. The reverse belief message from TX $j$ to RX $i$ is
denoted $p_{i \la j}(t,\cdot): \Xset \mapsto\R$. $p_{i \la
j}(t,\xbf_j)$ and $p_{i \la j}(t,\xbf_j)$ denote the values of the
beliefs at $\xbf_j$. After some fixed number of rounds, the
algorithm is stopped and a final scheduling decision, meaning a
selection of the TX vectors $\xbf_j$, is made by the transmitters.
The steps for BP are as follows:

\begin{enumerate}
\item \emph{Initialization:}  Set $t=0$ and for all $(i,j) \in E$, let
$p_{i \la j}(t,\xbf_j)$ be some initial distribution on $\xbf_j$.
This distribution could be, for example, the uniform distribution on
the set $\Xset$.
\item \emph{RX node update:}  In the RX half of the round, each RX $i$
sends a belief message to the transmitters TX $j$ with $j \in
\Nrx(i)$ given by \beq \label{eq:outBP}
    p_{i \ra j}(t,\xbf_j) =
        \Exp\left[ \exp(uf_i(\xbf_i,\zbf_i)) \mid \xbf_j \right],
\eeq where $\zbf_i = \Abf_i \xbf$ as in (\ref{eq:zidef}) and the
expectation is over independent $\xbf_k \sim p_{i \la k}(t,\xbf_k)$,
$\forall k\in \Nrx(i)$.
\item \emph{TX node update:}  In the TX half of the round, each TX $j$
sends a belief message back to the receivers RX $i$ with $i \in
\Ntx(j)$ given by \beq \label{eq:inBP}
    p_{i \la j}(t+1,\xbf_j) = \frac{1}{Z}
    \prod_{\ell \in \Ntx(j) \neq i} p_{\ell \ra j}(t,\xbf_j),
\eeq where $Z$ is a normalization constant and the product is over
all $\ell \in \Ntx(j)$ with $\ell \neq i$. The iteration number is
incremented, $t=t+1$, and we return to step 2 until a sufficient
number of rounds have been performed.
\item \emph{Final solution:}  The final estimate for the marginal
distribution of $\xbf_j$ is given by \beq \label{eq:xhatBP}
    p_j(t+1,\xbf_j) = \frac{1}{Z} \prod_{i \in \Ntx(j)} p_{i \ra j}(t,\xbf_j).
\eeq The scheduling vector can be selected as the maximum of this
marginal distribution.
\end{enumerate}

In the case when the graph $G$ has no cycles, it can be shown that
the $p_j(t,\xbf_j)$ converges to the true marginal distribution of
$p(\xbf)$ in (\ref{eq:probF}) with respect to the variable $\xbf_j$.
However, for general $G$, BP is approximate.  A complete treatment
of BP is beyond the scope of this work -- the reader is referred to
the references above.

Implementation of the above BP algorithm in wireless networks is
challenging due to two reasons:
\begin{itemize}
\item \emph{High computational complexity:} The
expectation in \eqref{eq:outBP} requires integration over all the
variables $\xbf_r$ with $r \in \Nrx(i)$ and $r \neq j$. This
computation grows exponentially in $|\Nrx(i)|$, which is the number
of transmitters interfering with RX $i$.  If this set is large, the
computation is prohibitive.
\item \emph{High messaging overhead:} Passing the beliefs requires
\emph{unicast} messages between each RX and each TX (which are
neighbors as per graph $G$) as opposed to single broadcast message.
Also, in each round $t$, the messages comprise of the beliefs $p_{i
\ra j}(t,\xbf_j)$ and $p_{i \la j}(t,\xbf_j)$ for all values $\xbf_j
\in \Xset$. If $\Xset$ is large, the messaging overhead may be
significant, and it grows with the number of transmitters
interfering at a receiver.
\end{itemize}
\subsection{Gaussian Approximation} \label{sec:gaussBP}

We first consider the simplification of the RX node update
\eqref{eq:outBP}. We describe the simplification in log domain. Let
$\Delta_{i \ra j}(t,\cdot)$ and $\Delta_{i \la j}(t,\cdot)$ be
\emph{log likelihood functions}, meaning any functions such that
\begin{subequations} \label{eq:DeltaLR}
\beqa
    \Delta_{i \ra j}(t,\xbf_j) &:=& \frac{1}{u}  \log \left[
        p_{i \ra j}(t,\xbf_j) \right] + \mbox{const}, \label{eq:DeltaRight} \\
    \Delta_{i \la j}(t,\xbf_j) &:=& \frac{1}{u}  \log \left[
        p_{i \la j}(t,\xbf_j) \right] + \mbox{const}. \label{eq:DeltaLeft}
\eeqa
\end{subequations}
where the constants do not depend on $\xbf_j$ (although they may
depend on $t$ and the indices $i$ and $j$). Observe that we can
recover the probabilities from the log likelihoods by the relation
\begin{subequations} \label{eq:probDeltaLR}
\beqa
    p_{i \ra j}(t,\xbf_j) &\propto& \exp\left[ u \Delta_{i \ra j}(t,\xbf_j) \right]
        \label{eq:probDeltaR} \\
    p_{i \la j}(t,\xbf_j) &\propto& \exp\left[ u \Delta_{i \la j}(t,\xbf_j) \right].
        \label{eq:probDeltaL}
\eeqa
\end{subequations}

We now consider the simplification of the RX node update
\eqref{eq:outBP} under two cases:  when $i=j$ and when $i \neq j$.
We begin with the case when $i=j$. The expectation in
(\ref{eq:outBP}) is to be evaluated with $\zbf_i$ given by
(\ref{eq:zidef}) with the variables $\xbf_j$ being independent and
$\xbf_j \sim p_{i \la j}(t,\xbf_j)$. Let $\xbfhat_{i \la j}(t)$ and
$\Qbf^x_{i \la j}(t)$ be the mean and $1/u$ times the variance of
the distribution $p_{i \la j}(t,\cdot)$. Then, under the simplifying
assumption that the summation in (\ref{eq:zidef}) consists of a
large number of independent terms, we can apply the Central Limit
Theorem and approximate the distribution of $\zbf_i$ with \beq
\label{eq:ziGauss}
    \zbf_i = {\cal N}(\sbfhat_{ii}(t), \Qbf^s_{ii}(t)/u),
\eeq where
\begin{subequations} \label{eq:supdate}
\beqa
    \sbfhat_{ii}(t) &=& \sum_{j \in \Nrx(i)} \Abf_{ij} \xbfhat_{i \la j}(t) \\
    \Qbf^s_{ii}(t) &=& \sum_{j \in \Nrx(i)} \Abf_{ij} \Qbf^x_{i\la j}(t)\Abf_{ij}',
\eeqa
\end{subequations}
which have the interpretation as a mean and variance of the
interference at RX $i$. Applying the Gaussian approximation
\eqref{eq:ziGauss} to the expectation (\ref{eq:outBP}) shows that we
can write $\Delta_{i \ra i}(\cdot)$ in \eqref{eq:DeltaRight} as \beq
\label{eq:DeltaiiRZ}
    \Delta_{i \ra i}(t,\xbf_i) \approx \frac{1}{u} \log \left[
        Z_{i0}(\xbf_i,\sbfhat_{ii}(t), \Qbf^s_{ii}(t)) \right],
\eeq where $Z_{i0}(\cdot)$ is the partition function \beq
\label{eq:Zi0}
    Z_{i0}(\xbf_i,\sbfhat_{ii},\Qbf^s_{ii}) = \int
        \exp\left[ uL_{i0}(\xbf_i,\zbf_i, \sbfhat_{ii},\Qbf^s_{ii})\right] d\zbf_i,
\eeq and \beq \label{eq:Li0}
    L_{i0}(\xbf_i,\zbf_i, \sbfhat_{ii},\Qbf^s_{ii}) = f_i(\xbf_i,\zbf_i)
    - \frac{1}{2} (\zbf_i - \sbfhat_{ii})'\Qbf^{-s}_{ii}(\zbf_i - \sbfhat_i)
\eeq and $\Qbf_{ii}^{-s}$ is the matrix inverse of $\Qbf_{ii}^s$.

Next consider the case $i \neq j$. A similar argument as above shows
that, conditional on $\xbf_j$, the distribution of $\zbf_i$  can be
approximated by the Gaussian \beq \label{eq:zijGauss}
    \zbf_i = {\cal N}(\sbfhat_{ij}(t), \Qbf^s_{ij}(t)/u),
\eeq where
\begin{subequations} \label{eq:supdateij}
\beqa
    \sbfhat_{ij}(t) &=& \sum_{r \in \Nrx(i) \neq j} \Abf_{ir} \xbfhat_{i \la r}(t)
        + \Abf_{ij}\xbf_j \nonumber \\
        &=& \sbfhat_{ii}(t) + \Abf_{ij}(\xbf_j-\xbfhat_{i \la j}(t)) \\
    \Qbf^s_{ij}(t) &=& \sum_{r \in \Nrx(i) \neq j} \Abf_{ir} \Qbf^x_{i\la r}(t)\Abf_{ir}', \nonumber \\
        &=& \Qbf^s_{ii}(t) - \Abf_{ij} \Qbf^x_{i\la j}(t)\Abf_{ij}'.
\eeqa
\end{subequations}
Then, applying the Gaussian approximation (\ref{eq:zijGauss}) along
with \eqref{eq:DeltaLeft} to the expectation (\ref{eq:outBP}) shows
that we can write $\Delta_{i \ra j}(\cdot)$ in \eqref{eq:DeltaRight}
as \beq \label{eq:DeltaijRZ}
    \Delta_{i \ra j}(t,\xbf_j)
    \approx \frac{1}{u}\log\left[ Z_i(\Delta_{i \la i}(t,\cdot),\sbfhat_{ij}(t), \Qbf^s_{ij}(t))
        \right],
\eeq where
\[
    Z_{i}(\Delta(\cdot),\sbfhat_{ij},\Qbf^s_{ij}) = \int
        \exp\left[ uL_{i}(\xbf_i,\zbf_i, \sbfhat_{ij},\Qbf^s_{ij})\right] d\xbf_i d\zbf_i,
\]
and \beq \label{eq:Li}
    L_{i}(\xbf_i,\zbf_i, \sbfhat_{ij},\Qbf^s_{ij}) = L_{i0}(\xbf_i,\zbf_i, \sbfhat_{ij},\Qbf^s_{ij}) + \Delta(\xbf_i).
\eeq

Then, the RX and TX node update steps of the BP algorithm with
Gaussian approximation of the interference at the RX are:
\begin{itemize}
\item \emph{RX node update:}
The above equations are used to simplify the standard BP algorithm
as follows. Each RX $i$ receives mean and variances $\xbfhat_{i \la
j}(t)$ and $\Qbf^x_{i \la j}(t)$ of the vectors from the
transmitters TX $j$ with $j \in \Nrx(i)$. RX $i$ also receives the
function $\Delta_{i \la i}(t,\cdot)$ from its serving transmitter,
TX $i$. RX $i$ then computes the interference means and variances in
\eqref{eq:supdate} and \eqref{eq:supdateij}. Then, for each TX $j$
it sends back the log likelihoods $\Delta_{i \ra j}(t,\cdot)$ by
evaluating the log partition functions \eqref{eq:DeltaiiRZ} and
\eqref{eq:DeltaijRZ}.

\item \emph{TX node update:}
It can be verified that converting the update \eqref{eq:inBP} to log
domain yields \beq \label{eq:DeltaijLSum}
    \Delta_{i \la j}(t+1,\xbf_j) = \sum_{\ell \in \Ntx(j) \neq i}
        \Delta_{\ell \ra j}(t,\xbf_j).
\eeq Each TX $j$ can first computes the log likelihoods $\Delta_{i
\la j}(t+1,\xbf_j)$ from the log likelihoods $\Delta_{\ell \ra
j}(t,\xbf_j)$ from the receivers RX $\ell$ with $\ell \in \Ntx(j)$,
$\ell \neq i$. Then, using the log likelihoods $\Delta_{i \la
j}(t+1,\xbf_j)$ , TX $j$ computes the mean and variance $\xbfhat_{i
\la j}(t+1)$ and $\Qbf^x_{i \la j}(t+1)$ from the probability
distribution $p_{i \la j}(t+1,\xbf_j)$ in \eqref{eq:probDeltaL}.
Then, TX $j$ sends messages to the receivers as described in the RX
node update above.
\end{itemize}
After a sufficient number of rounds, TX $j$ can compute the final
log likelihood \beq \label{eq:DeltajLSum}
    \Delta_{j}(t+1,\xbf_j) = \sum_{\ell \in \Ntx(j)} \Delta_{\ell \ra j}(t,\xbf_j),
\eeq and then compute the final scheduling vector as the one which
maximizes the above function.

We have therefore simplified the standard BP algorithm by
eliminating the exponential complexity of the RX update
\eqref{eq:outBP}, and replaced this computation with a Gaussian
approximation.

\subsection{First Order Approximations} \label{sec:firstOrderApprox}
Unfortunately, the Gaussian approximation above does not
significantly reduce the messaging overhead.  Each RX and TX must
still send separate unicast messages to every TX or RX in its
neighbor set.  Also, although the TX must only send a mean and a
covariance matrix $\xbfhat_{i \la j}(t)$ and $\Qbf^x_{i \la j}(t)$,
the receivers must send the entire log likelihood functions
$\Delta_{i \ra j}(t,\xbf_j)$.

The messaging overhead can be reduced via selective use of first
order approximations as follows: Divide the edges $(i,j) \in E$ with
$i\neq j$
 into two sets -- \emph{weak} and \emph{strong} -- depending on whether
 $\Abf_{ij}$ is small or large.
Along a strong edge $(i,j)$, RX $i$ and TX $j$ exchange the full
unicast messages described above. However, for the weak edges, the
messages can be replaced with a first order approximation described
below. The precise classification rule between weak and strong edges
is an algorithm parameter that can be used to trade off complexity
and accuracy.

To describe the first order approximation along the weak edges,
suppose $\Abf_{ij}$ is small for some edge $(i,j)$. Let
$\hat{\xbf}_j(t)$ be the mean value of the distribution
corresponding to the log likelihood $\Delta_j(t,\cdot)$ in
\eqref{eq:DeltajLSum}. Now consider the log likelihood $\Delta_{i
\ra j}(t,\xbf_j)$ in \eqref{eq:DeltaijRZ}. Applying
\eqref{eq:supdateij}
\beqa
    \lefteqn{ \Delta_{i \ra j}(t,\xbf_j)
    \approx \frac{1}{u}\log\left[ Z_i(\Delta_{i \la i}(t,\cdot),\sbfhat_{ij}(t), \Qbf^s_{ij}(t))
        \right]}, \nonumber \\
        &\stackrel{(a)}{\approx}&
        \frac{1}{u}\log\left[ Z_i(\Delta_{i \la i}(t,\cdot),\right. \nonumber\\
        & &\left. \sbfhat_{ii}(t)+
            \Abf_{ij}(\xbf_j-\xbfhat_{j}(t)), \Qbf^s_{ii}(t)) \right], \nonumber \\
       &\stackrel{(b)}{\approx}&
            \ubf'_{ij}(t) \xbf_j  + \mbox{ const}     \label{eq:DelijApprox}
\eeqa
where (a) follows from \eqref{eq:supdateij} and the
approximation that $\Qbf^s_{ij}(t) \approx \Qbf^s_{ii}(t)$ and
$\xbfhat_{i \la j}(t) \approx \xbfhat_j(t)$ when $\Abf_{ij}$ is
small; and (b) follows from taking a Taylor's approximation with
\beq \label{eq:uij}
    \ubf_{ij}(t) = \Abf'_{ij}D'_{i1}(t)-
            \Abf_{ij}'D_{i2}(t)\Abf_{ij}\xbfhat_{j}(t),
\eeq
where for $r=1,2$, $D_{ir}(t)$, are the derivatives
\beq \label{eq:logZDeriv}
    D_{ir}(t) := \frac{1}{u}\frac{\partial^r}{\partial \sbfhat^r}
        \log \left[Z_i(\Delta_{i \la i}(\cdot),\sbfhat_{ii}(t), \Qbf^s_{ii}(t)) \right].
\eeq
The constant in \eqref{eq:DelijApprox} is independent of
$\xbf_j$. Using standard properties of the cumulant function
\cite{WainwrightJ:08}, it can be shown that the derivative is given
by
\beq \label{eq:derivUtil}
    D_{ir}(t) = \Exp\left[ \left.
    \frac{\partial^r}{\partial \zbf_i^r} f_i(\xbf_i,\zbf_i)
    \right| \sbfhat_{ii}(t),\Qbf^s_{ii}(t), \Delta_{i \la i}(t,\cdot) \right],
\eeq
where the expectation is with respect to the conditional distribution
\beqa
    \lefteqn{ p_{\xbf,\zbf|\sbfhat,\Qbf^s}(\xbf_i,\zbf_i\mid
    \sbfhat_{ii}(t),\Qbf^s_{ii}(t)) }     \nonumber \\
     &:=& \frac{1}{Z_i}\exp\left[
        uL_i(\xbf_i,\zbf_i, \sbfhat_{ii}(t),\Qbf^s_{ii}(t)\right],
        \label{eq:pxz}
\eeqa
and $L_i(\cdot)$ is defined in \eqref{eq:Li} and
\eqref{eq:Li0}. The dependence on $\Delta_{i \la i}(\cdot)$ in \eqref{eq:derivUtil}
is implicit in \eqref{eq:Li}.
Note that the derivatives $D_{ir}(t)$ in
\eqref{eq:derivUtil} can be interpreted as a sensitivity of the
expected utility $f_i(\xbf_i,\zbf_i)$ to changes in the interference
$\zbf_i$.

The computation of $\xbf_{i \la j}(t)$ can also be simplified as
follows: Recall that $\xbfhat_{i \la j}(t)$ and $\xbfhat_j(t)$ are
the expectation with respect to the likelihood functions $\Delta_{i
\la j}(t,\cdot)$ in \eqref{eq:DeltaijLSum} and $\Delta_j(t,\cdot)$
in \eqref{eq:DeltajLSum}, respectively. Therefore, we can write them
as
\begin{subequations}
\beqa
    \xbfhat_{i \la j}(t) &=& \Exp\left[ \xbf_j | \Delta_{i \la j}(t,\cdot) \right]
        \label{eq:xijDel} \\
    \xbfhat_{j}(t) &=& \Exp\left[ \xbf_j | \Delta_{j}(t,\cdot) \right],  \label{eq:xjDel}
\eeqa
\end{subequations}
where we have used the notation $\Exp( g(\xbf) | \Delta(\xbf) )$ to
denote the expectation of $g(\xbf)$ with respect to the probability
distribution
\[
    p_\Delta(\xbf) \propto \exp\left[u \Delta(\xbf) \right].
\]
It is easy to check that for small perturbations of $\epsilon(\xbf)$
of $\Delta(\xbf)$, we have the first order approximation \beqa
    \lefteqn{ \Exp(g(\xbf)|\Delta + \epsilon) \approx
        \Exp(g(\xbf)|\Delta)   } \nonumber \\
    &+& u\left[ \Exp(g(\xbf)\epsilon(\xbf)|\Delta) -
    \Exp(g(\xbf)|\Delta)\Exp(\epsilon(\xbf)|\Delta)\right]. \label{eq:EdelApprox}
\eeqa
Therefore,
\beqa
    \lefteqn{ \xbfhat_{i \la j}(t+1) \stackrel{(a)}{=}
        \Exp\left[ \xbf_j | \Delta_j(t+1,\cdot) - \Delta_{i \ra j}(t,\cdot)\right]
        } \nonumber \\
    &\stackrel{(b)}{\approx}&
        \Exp(\xbf_j|\Delta_j(t+1,\cdot)) -
        u\Exp(\xbf_j\Delta_{i \ra j}(t,\xbf_j)|\Delta_j(t+1,\cdot)) \nonumber \\
    &-& u\Exp(\xbf_j|\Delta_j(t,\cdot))
     \Exp(\Delta_{i \ra j}(t,\xbf_j)|\Delta_j(t+1,\cdot)) \nonumber \\
    &\stackrel{(c)}{\approx}&  \xbfhat_j(t+1) - \Qbf^x_j(t+1)\ubf_{ij}(t)
    \label{eq:xijApprox}
\eeqa
where (a) follows from \eqref{eq:xijDel} along with
\eqref{eq:DeltaijLSum} and \eqref{eq:DeltajLSum}; (b) follows from
\eqref{eq:EdelApprox}; and (c) follows from \eqref{eq:xjDel} and
\eqref{eq:DelijApprox}.

We can use the above relations to define the following algorithm:
\begin{enumerate}
\item \emph{Initialization:}  Set $t=0$.  Each TX $j$
broadcasts an initial scheduling vector $\xbfhat_j(t)$ and variance
$\Qbf^x_j(t)$. These can be based on the mean and variance of
$\xbf_j$ over the set $\Xset$. The receivers RX $i$ sets $\xbfhat_{i
\la j}(t) = \xbf_j(t)$ and $\Qbf^x_{i\la j}(t)=\Qbf^x_j(t)$ for all
$j$, and $\Delta_{i \la i}(t,\xbf_i) = 0$ for all $\xbf_i$.

\item \emph{RX node update:}  In the RX half of the round,
each RX $i$ first computes the interference means and variances
$\sbfhat_{ii}(t)$ and $\Qbf^s_{ii}(t)$ in \eqref{eq:supdate} and
$\sbfhat_{ij}(t)$ and $\Qbf^s_{ij}(t)$ in \eqref{eq:supdateij}.
Then, RX $i$ computes the log likelihood function $\Delta_{i \ra
i}(t,\xbf_i)$ in \eqref{eq:DeltaiiRZ} and sends it as a unicast
message to its serving transmitter TX $i$. Also, for each strong
edge $(i,j)$, RX $i$ computes the log likelihood functions
$\Delta_{i \ra j}(t,\xbf_j)$ in \eqref{eq:DeltaijRZ} and sends it as
a unicast message to the interfering TX $j$.  For the weak edges, RX
$i$ simply computes the sensitivity $D_{i1}(t)$ in
\eqref{eq:derivUtil} and broadcasts it to all other transmitters TX
$j$. The receiver also computes $\ubf_{ij}(t)$ in \eqref{eq:uij} and
stores it for the next round.

\item \emph{TX node update:}  In the TX half of the round, each transmitter TX
$j$ would have received the log likelihoods $\Delta_{i \ra
j}(t,\xbf_j)$ from the receivers RX $i$, along the edges $(i,j)$
that were strong.  For any weak edge $(i,j)$, TX $j$ can
approximately compute the log likelihood from the sensitivity
$D_{i1}(t)$ using \eqref{eq:DelijApprox}. TX $j$ can then compute
the log likelihoods $\Delta_{i \la j}(t+1,\xbf_j)$ from
\eqref{eq:DeltaijLSum} for all $i \in \Ntx(j)$ and the log
likelihood $\Delta_j(t+1,\xbf_j)$ in \eqref{eq:DeltajLSum}. TX $j$
sends the receiver RX $j$ that it is serving the entire log
likelihood $\Delta_{j \la j}(t+1,\xbf_j)$. For the receivers RX $i$
such that $(i,j)$ is a strong edge, TX $j$ computes the mean
$\xbfhat_{i \la j}(t+1)$ and variance $\Qbf_{i \la j}^x(t+1)$ from
the log likelihood $\Delta_{i \la j}(t+1,\xbf_j)$ and sends it as a
unicast message to RX $i$. Each TX $j$ also computes the mean and
variance $\xbfhat_j(t+1)$ and $\Qbf^x_j(t+1)$ from the log
likelihood $\Delta_j(t+1,\xbf_j)$ and broadcasts the quantities to
the other receivers.  Any receiver RX $i$ that is along a weak edge
$(i,j)$ can then approximately compute $\xbfhat_{i \la j}(t)$ from
\eqref{eq:xijApprox}.

The round number is incremented, $t=t+1$, and we return to Step 2
until a fixed number of rounds have been performed.

\item \emph{Final solution:}  After the final round,
each transmitter TX $j$ takes the scheduling vectors to be the
vector $\xbf_j$ that maximizes the log likelihood
$\Delta_j(t,\xbf_j)$.
\end{enumerate}

\begin{figure}
\begin{center}
  \includegraphics[height=2in]{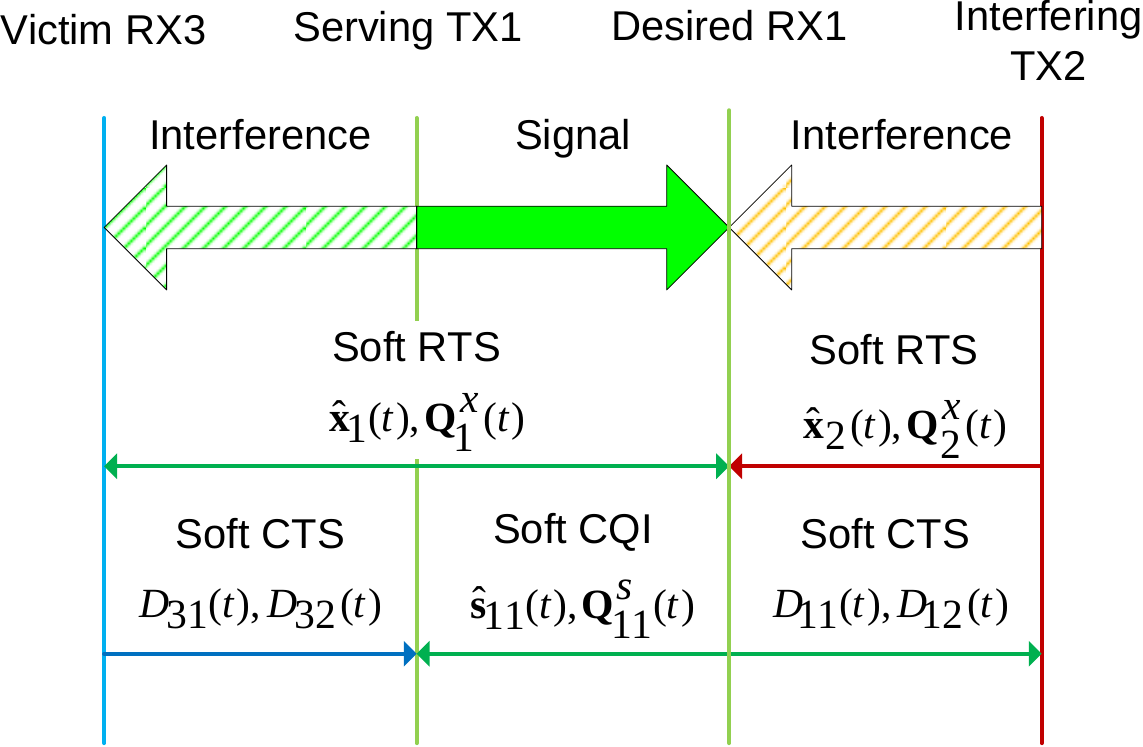}
\end{center}
\caption{One round of the BP messages along the weak edges
interpreted as a ``soft" RTS / CTS mechanism.
}
\label{fig:bpMsgFlow}
\end{figure}

The message flow along the weak edges has an appealing
interpretation. Consider Fig.\ \ref{fig:bpMsgFlow} where a
transmitter TX1 attempts to send data to a receiver RX1.  The
receiver RX1 experiences interference from a second transmitter TX2,
while the transmitter TX1 causes interference onto a victim receiver
RX3.  Fig.\ \ref{fig:bpMsgFlow} shows the messages along the weak
edges in one round of the BP algorithm to coordinate the
interference. The transmitters TX1 and TX2 will broadcast the mean
and variance $\xbfhat_j(t)$ and $\Qbf^x_j(t)$ of their intended
transmit vectors. These transmissions can be interpreted as ``soft"
request to sends (RTS).  They are soft since the intended transmit
vectors are signaled by a distribution.  Based on the transmit
vector distribution from the interfering TX2, the
 receiver RX1 replies to the serving TX1 with an estimate of the interference
 $\sbfhat_{11}(t)$ and $\Qbf^s_{11}(t)$ which can be interpreted
as a ``soft" channel
quality indication (CQI).  As victim receivers, RX1 and RX3 also compute
the sensitivities to the interference level by the derivatives $D_{ir}(t)$.
These values can be interpreted as soft clear to send (CTS) indications,
since instead of a binary go/no go type CTS, they signal a soft cost
on changes in the interference from other transmitters.
\subsection{Communication and Computation Complexity}
We summarize the complexity of one round of the three variations of
BP in Tables~\ref{tab:comp_compl} and~\ref{tab:comm_compl}.
Specifically, we focus on RX $i$ and TX $j$ and consider how the
complexity grows with $\Xset$, $\Nrx(i)$ and $\Ntx(j)$.

\begin{table}
\begin{center}
\begin{tabular}{|c|c|c|}
 \hline   Method & RX $i$ & TX $j$ \\
\hline Exact & $O\left(\left|\Xset\right|^{|\Nrx(i)|}\right)$&  $O\left(|\Xset||\Ntx(j)|\right)$ \\
\hline Gaussian Approx. & $O\left( |\Nrx(i)| \right)$ & $O\left(|\Xset||\Ntx(j)|\right)$\\
\hline First Order &$O\left(|\Nrx(i)| \right)$ & $O\left(|\Ntx(j)| \right)$\\
\hline
\end{tabular}
\end{center}
\caption{Computational Complexity}
\label{tab:comp_compl}
\end{table}

\begin{table}
\begin{center}
\begin{tabular}{|c|c|c|}
 \hline Method & RX $i$ & TX $j$ \\
\hline Exact & $O\left(|\Xset||\Nrx(i)|\right)$& $O\left(|\Xset||\Ntx(j)|\right)$\\
\hline Gaussian Approx. & $O\left(|\Xset||\Nrx(i)|\right)$&  $O\left(|\Ntx(j)|\right)$\\
\hline First Order & O(1) & O(1)\\
\hline
\end{tabular}
\end{center}
\caption{Communication Complexity}
\label{tab:comm_compl}
\end{table}

\section{Numerical Simulation} \label{sec:sim}

The BP algorithm was simulated on a
a simplified version of an industry standard
model for LTE femtocell evaluation in \cite{FemtoForum:10}.
The simulation parameters are shown in Table \ref{tbl:simParam}.
The network consists of a 3 x 3 grid of 10m x 10m apartments
with active links in 5 of the 9 apartments.
Each link consists of one femto BS transmitting to one femto mobile
(called a UE, or user equipment, in 3GPP terminology).
Due to restricted association, UEs connect to the femto BS in their
apartment even if it is not the BS with the minimum path loss.
As mentioned in the introduction, this scenario exposes many links
to strong interference, and thus presents a good test scenario for
advanced interference coordination algorithms.

\begin{table}
\begin{center}
\begin{tabular}{|p{1in}|p{2in}|}
\hline
Parameter & Value \\ \hline
Network topology & $3 \x 3$ apartment model, with active links in 5 of the 9 apartments.
\\ \hline
Carrier freq & 2 GHz \\ \hline
Bandwidth & 5 MHz \\ \hline
Wall loss & 0 or 10 dB \\ \hline
Lognormal shadowing & 10 dB std.\ dev.\ \\ \hline
Path loss & $38.46 + 20\log_{10}(R) + 0.7R$ dB, $R$ distance in meters. \\ \hline
Femto BS TX power & 0 dBm \\ \hline
Femto UE noise figure & 4 dB \\ \hline
\end{tabular}
\end{center}
\caption{Simulation parameters.}
\label{tbl:simParam}
\end{table}

In the first simulation, we considered a time-varying simulation
with a simple on-off model where, in each time slot,
each link either transmits at the max power or is completely off.
As described in Section \ref{sec:ex}, for the time-varying problem,
the utility maximization was rerun in each
time slot with the weighted sum rate utility \eqref{eq:wtUtil} with weights
\eqref{eq:wtMargUtil}.  We used the proportional fair utility $U_i(R) = \log(R)$.
We generated independent flat fades on the links in each slot,
and took a filter time constant in \eqref{eq:rateFilt} was $\alpha = 0.1$.
For each random realization, or ``drop", of the femto network, we ran the simulation
over 100 time slots and measured the average rate.
The wall loss in the femto model was 10 dB.

\begin{figure}
\begin{center}
  \includegraphics[height=2.5in]{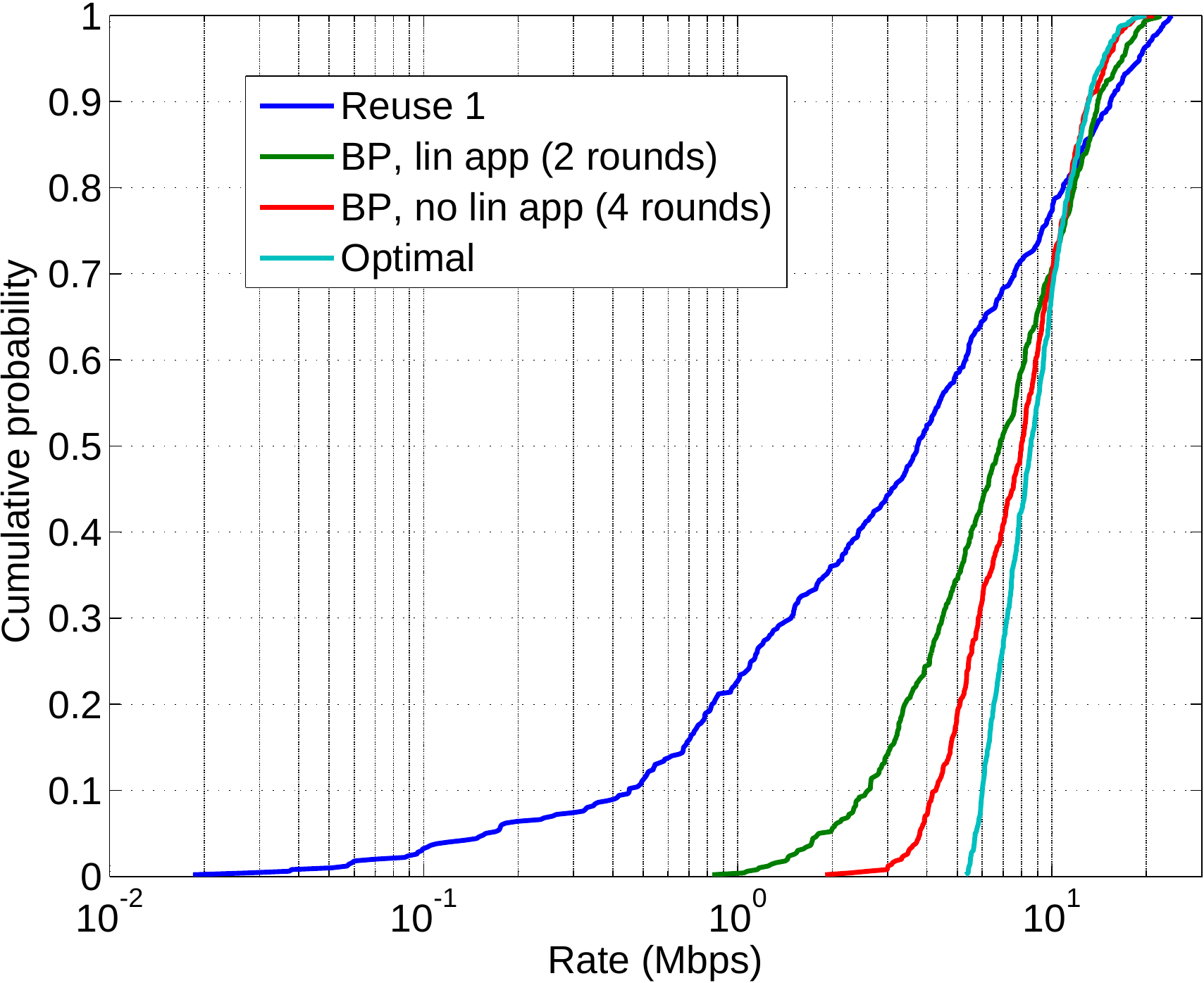}
  \includegraphics[height=2.5in]{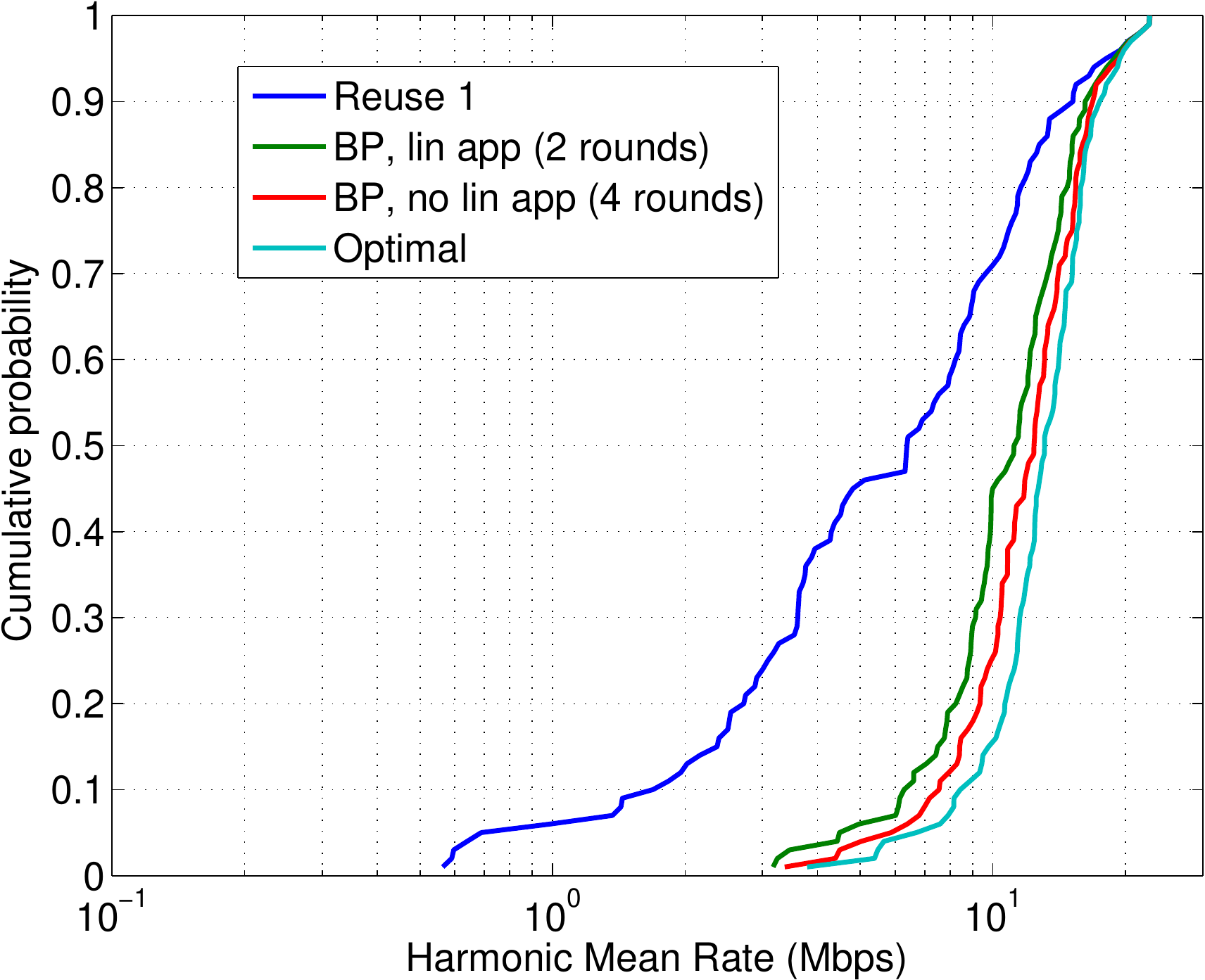}
\end{center}
\caption{Downlink femtocell simulation with an on-off channel model
and time-varying flat fading.
The top panel shows the rate CDFs across all links and drops
using various optimization algorithms for weight matching.
The bottom panel is the CDF of system utility (represented
as the harmonic mean rate) across different drops.}
\label{fig:dynSim}
\end{figure}

The top panel of Fig.\ \ref{fig:dynSim} plots
the cumulative distribution function (CDF)
of the time-averaged rates for 5 links and 100 drops
comparing various optimization methods for computing the maximum weighted
matching optimization \eqref{eq:opt}.
The curve ``reuse 1" is the case when all links transmit at max power.
Two cases of BP
are simulated:  (i) 4 rounds of BP in each time slot
using the Gaussian approximation in Section
\ref{sec:gaussBP} but no first order approximations; and
(ii) using only two rounds of BP
with both the Gaussian approximations and first order approximations on all interfering links less than 0 dB below the serving link.
We see that even the linear approximated BP with 2 rounds does significantly better
than reuse 1, and is not that far from the 4 round BP.
The gap between BP and reuse 1 particularly large at low rates.
For example, for the 20\% worst links,
BP offers almost a factor of 5 improvement in rate over reuse 1.

Also plotted is the rate CDF with optimal matching in each time slot
based on an exhaustive centralized search.
BP performs reasonably close to this curve, although there is still an
obvious gap, again at low rates.

The bottom panel of Fig.\ \ref{fig:dynSim} shows the CDF
of the total system utility over the 100 drops.  Since the optimization used
a log utility, the optimization is equivalent to maximizing the harmonic
mean rate over the 5 links.  We see in this plot that the approximate BP
algorithms are close to optimal and significantly better than reuse 1.

As a second simulation, we considered the same problem but with a
static single optimization and $K=4$ subbands.  Random fading was
generated in each subband, so the simulation would be applicable
in the case with frequency selective fading with coherence bandwidth
roughly equal to the subband bandwidth.  The optimization was over
 scheduling vectors  $\xbf_j \in \C^K$ with each component being on or off,
 so there are $2^K-1=15$ possible non-zero scheduling vectors in each link.
Optimal subband allocation is a well-known
challenging but important problem for OFDMA systems like LTE.

\begin{figure}
\begin{center}
  \includegraphics[height=2.5in]{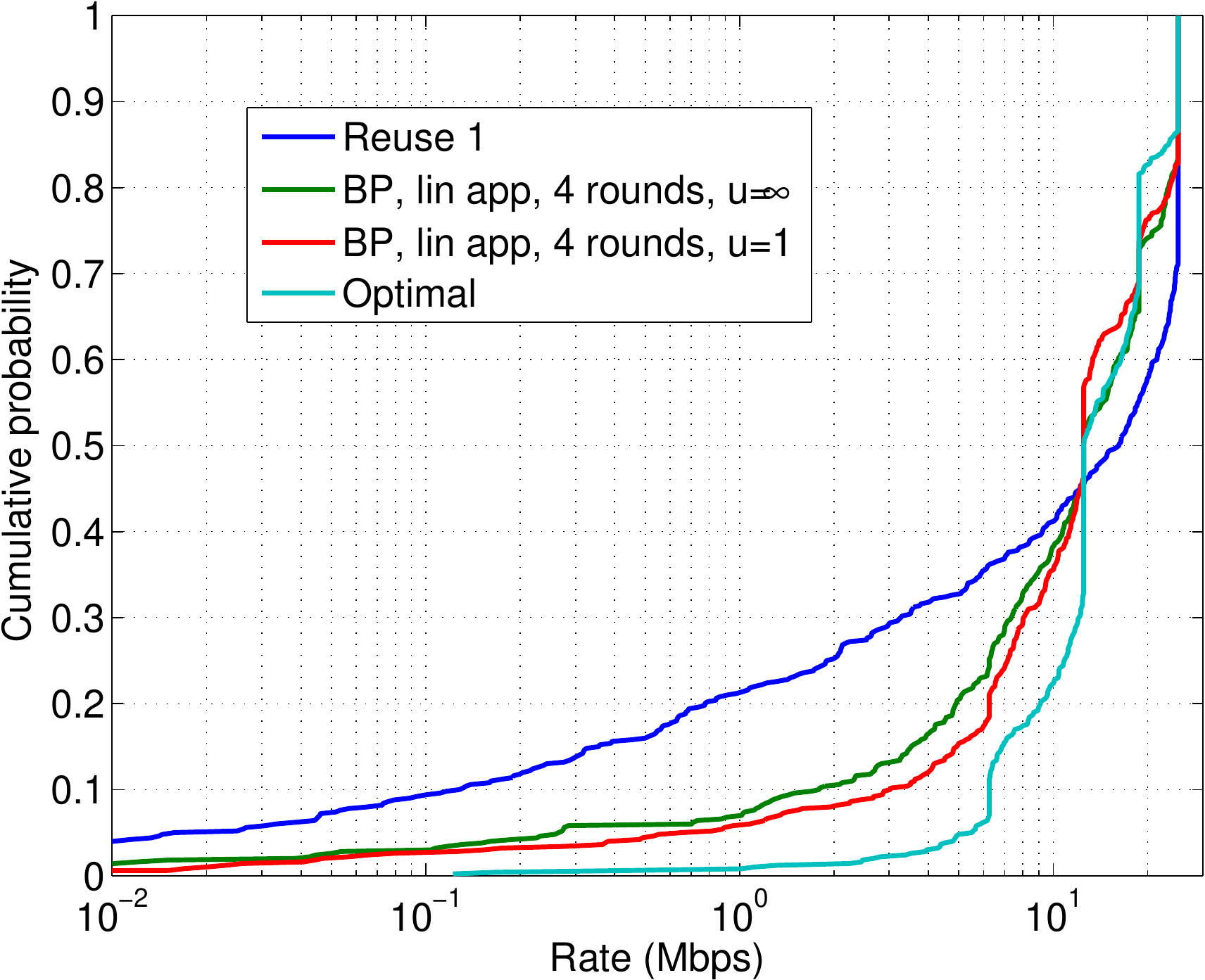}
\end{center}
\caption{Subband static optimization.
Performance comparison for downlink femtocell rates with
various optimizations for 4 independently faded subbands.
}
\label{fig:subSim}
\end{figure}

Under these assumptions, Fig.\ \ref{fig:subSim} shows the CDF
of the rates under various optimization methods.
Similar to the dynamic single subband case, we see that BP
provides significant gains over simple reuse 1, even when we use
linear approximations and only four rounds.  Also, for most links,
BP achieves a rate reasonably close to the optimal subband allocation found by
exhaustive search over all subband allocations over all rates.
However, for the lowest rate links, there is a significant gap
between BP and optimal.  Thus, even though BP outperforms reuse 1
significantly in this regime, there is significant room for improvement.

\begin{figure}
\begin{center}
  \includegraphics[height=2.5in]{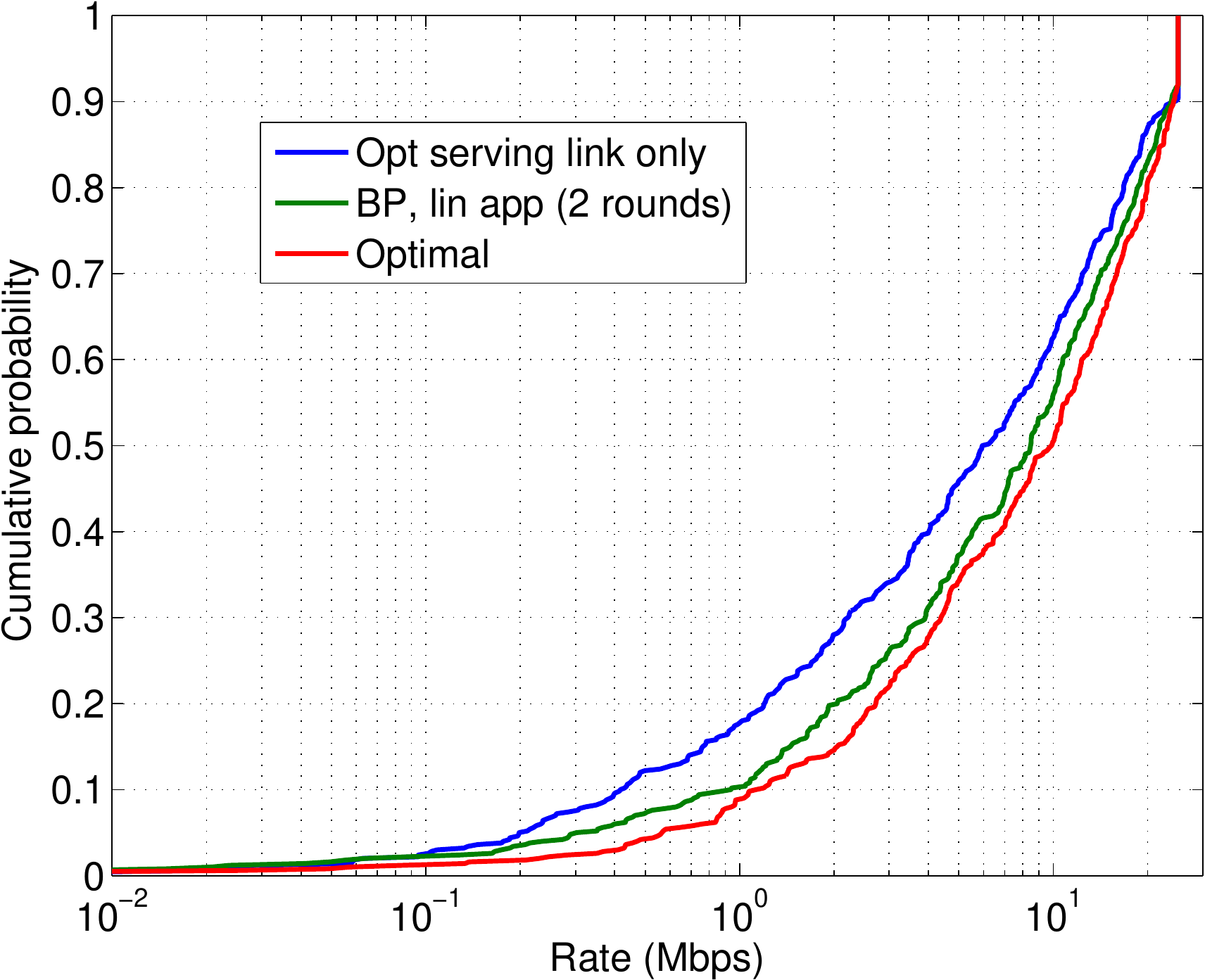}
\end{center}
\caption{Beamforming optimization.
Downlink femtocell rates for
various optimizations with 2 TX antenna beamforming.
Channel models assume no scattering and beamforming optimization
is performed over linear phase beamforming vectors.
}
\label{fig:bfSim}
\end{figure}

The BP method can also be applied to beamforming (BF) problems.
As a simple simulation, we consider again the femtocell deployment
with transmit beamforming with two antennas with half wavelength spacing
spacing and one receive antenna.  We neglect scattering so the channel
appears as a linear phase across the two antennas on all links.
For the transmit vectors $\xbf_j$, we optimize beamforming angles
over 10 angles uniformly spaced between 0 and $\pi$.
A performance comparison of the rate CDFs under various optimization
algorithms is shown in Fig.~\ref{fig:bfSim}.
In this case, we took a wall loss of 0 dB.
The curve labeled ``opt serving link only" is the case when the
BF vector is chosen to maximize the signal strength from the serving
link only without regard to interference.  This simple method provides
the baseline.  We see that BP provides some gains over serving link
optimization.  For example, the median rate with BP is approximately 50\%
higher than using optimal BF on the serving link only.
Also, BP appears to be reasonably close to the optimal BF selection
based on exhaustive search.

\section{Conclusions} \label{sec:conclusions}

We formulated a general wireless scheduling and interference
coordination problem as an optimization problem with linear mixing
utilities. This was cast in a BP framework where the goal is to
compute the marginal distributions of a joint probability function.
Using Gaussian and linear approximations, we obtained a distributed
interference coordination algorithms with low overhead. The
algorithm has a natural interpretation as a soft RTS/CTS scheme.
Numerical simulations demonstrated that the resulting algorithm is
close to an optimal scheme for dynamic and static sub-band
optimization as well as for beamforming coordination across cells.
Moreover, the results show that the algorithm computes a good
operating point in just two to four iterations making it very
attractive to be used in practical wireless cellular systems. In the
future, we plan to explore connections between our work and the AMP
framework in~\cite{MontanariT:06} to obtain performance bounds for
large random networks.

\bibliographystyle{IEEEtran}
\bibliography{bibl}

\end{document}